

\newcount\dflag
\dflag = 0


\def\monthname{\ifcase\month
\or Jan\ \or Feb\ \or Mar\ \or Apr\ \or May\ \or June\ %
\or July\ \or Aug\ \or Sept\ \or Oct\ \or Nov\ \or Dec\ 
\fi}




\def\endignore{}
\def\ignore #1\endignore{}


\global\hsize = 6.1in
\global\hoffset = .2in
\global\baselineskip = 1.2\baselineskip
\global\parskip = 4pt plus 0.3pt
\global\nulldelimiterspace = 0pt

\predisplaypenalty 5000


\overfullrule=0pt%

\font\titlerm = cmr10 scaled\magstep 4
\font\titlerms = cmr7 scaled\magstep 4
\font\titlermss = cmr5 scaled\magstep 4
\font\titlei = cmmi10 scaled\magstep 4
\font\titleis = cmmi7 scaled\magstep 4
\font\titleiss = cmmi5 scaled\magstep 4
\font\titlesy = cmsy10 scaled\magstep 4
\font\titlesys = cmsy7 scaled\magstep 4
\font\titlesyss = cmsy5 scaled\magstep 4
\font\titleit = cmti10 scaled\magstep 4

\def\titlefont{\def\rm{\fam0\titlerm}
\def\it{\fam\itfam\titleit}
\textfont0 = \titlerm
\scriptfont0 = \titlerms
\scriptscriptfont0 = \titlermss
\textfont1 = \titlei
\scriptfont1 = \titleis
\scriptscriptfont1 = \titleiss
\textfont2 = \titlesy
\scriptfont2 = \titlesys
\scriptscriptfont2 = \titlesyss
\textfont\itfam = \titleit
\rm}

\def\sectionfont{\def\rm{\fam0\tenrm}
\def\it{\fam\itfam\tenit}
\def\bf{\fam\bffam\tenbf}
\textfont0 = \tenrm
\scriptfont0 = \sevenrm
\scriptscriptfont0 = \fiverm
\textfont1 = \teni
\scriptfont1 = \seveni  \scriptscriptfont1=\fivei
\textfont2 = \tensy
\scriptfont2 = \sevensy
\scriptscriptfont2 = \fivesy
\textfont\itfam = \tenit
\textfont\bffam = \tenbf
\rm}

\font\littlefont = cmr7


\def\endid{}
\def\id#1\endid{\number\day\ \monthname \number\year
\hfill #1}

\def\endtitle{}
\def\title#1\endtitle{\vskip.15in\titlefont
\baselineskip = 1\baselineskip
#1\vskip0in
\baselineskip = 0.5\baselineskip\sectionfont
\global\baselineskip = 2 \baselineskip
}


\def\endauthors{}
\def\authors#1\endauthors{
#1}

\def\endabstract{}
\def\abstract#1\endabstract{\vskip .2in%
\centerline{\sectionfont\bf Abstract}%
\vskip .1in%
\noindent#1%
\footline = {\hfil}\pageno = 0
\vfill\eject
\pageno = 1\footline{\centerline{\sectionfont\folio}}
}


\newcount\nsection
\newcount\nsubsection

\def\section#1{\global\advance\nsection by 1
\global\nsubsection = 0
\def\eqlabel{(\number\nsection.1)}
\global\nequation = 1
\bigbreak\noindent
\centerline{\sectionfont\bf\number\nsection.\ #1}
\nobreak\smallskip\sectionfont\nobreak\par\nobreak}

\def\subsection#1{\global\advance\nsubsection by 1
\bigbreak\noindent
\centerline{\sectionfont \it \number\nsection.\number\nsubsection.\ #1}%
\nobreak\smallskip\rm\nobreak\par\nobreak}

\def\appendix#1#2{\bigbreak\noindent%
\centerline{\sectionfont \bf Appendix #1.\ #2}
\nobreak\smallskip\rm\nobreak\par\nobreak}


\newcount\nref
\global\nref = 1

\def\ref#1#2{\xdef #1{[\number\nref]}
#1
\ifnum\nref = 1\global\xdef\therefs{\noindent[\number\nref] #2\ }
\else
\global\xdef\oldrefs{\therefs}
\global\xdef\therefs{\oldrefs\vskip.1in\noindent[\number\nref] #2\ }%
\fi%
\global\advance\nref by 1
}

\def\aref#1{[\number\nref]
\ifnum\nref = 1\global\xdef\therefs{\noindent[\number\nref] #1\ }
\else
\global\xdef\oldrefs{\therefs}
\global\xdef\therefs{\oldrefs\vskip.1in\noindent[\number\nref] #1\ }%
\fi%
\global\advance\nref by 1
}

\def\listrefs{\vfill\eject\section{References}\therefs}


\newcount\cflag
\newcount\nequation
\global\nequation = 1

\def\nexteqno{\ifnum\cflag = 0
\global\advance\nequation by 1
\fi
\global\cflag = 0
\xdef\eqlabel{(\number\nsection.\number\nequation)}}

\def\lasteqno{\global\advance\nequation by -1
\xdef\eqlabel{(\number\nsection.\number\nequation)}}

\def\label#1{\xdef #1{(\number\nsection.\number\nequation)}
\ifnum\dflag = 1
{\escapechar = -1
\xdef\draftname{\littlefont\string#1}}
\fi}

\def\clabel#1#2{\xdef\eqlabel{(\number\nsection.\number\nequation #2)}
\global\cflag = 1
\xdef #1{\eqlabel}
\ifnum\dflag = 1
{\escapechar = -1
\xdef\draftname{\string#1}}
\fi}

\def\cclabel#1#2{\xdef\eqlabel{#2)}
\global\cflag = 1
\xdef #1{\eqlabel}
\ifnum\dflag = 1
{\escapechar = -1
\xdef\draftname{\string#1}}
\fi}


\def\eeq{}

\def\eqnn #1\eeq{$$ #1 $$}

\def\eq #1\eeq{\xdef\draftname{\ }
$$ #1
\eqno{\eqlabel \rlap{\ \draftname}} $$
\nexteqno}



\def\eol{& \eqlabel \rlap{\ \draftname} \crcr
\nexteqno
\xdef\draftname{\ }}

\def\eeol{& \eqlabel \rlap{\ \draftname}
\nexteqno
\xdef\draftname{\ }}

\def\eolnn{\cr
\global\cflag = 0
\xdef\draftname{\ }}


\def\eqa #1\eeq{\xdef\draftname{\ }
$$ \eqalignno{ #1 } $$
\global\cflag = 0}


\newcount\nfig
\global\nfig = 1

\def\fg#1\efig{\vskip .5in\noindent Fig.\ \number\nfig:\ #1%
\global\advance\nfig by 1}



\def\eg{{\it e.g.\/}}



\def\jref#1#2#3#4{{\it #1} {\bf #2}, #3 (#4)}

\def\NPB#1#2#3{\jref{Nucl.\ Phys.}{B#1}{#2}{#3}}

\def\PLB#1#2#3{\jref{Phys.\ Lett.}{#1B}{#2}{#3}}

\def\PRD#1#2#3{\jref{Phys.\ Rev.}{D#1}{#2}{#3}}

\def\PRV#1#2#3{\jref{Phys.\ Rev.}{#1}{#2}{#3}}
\def\PTP#1#2#3{\jref{Prog.\ Theor.\ Phys.}{#1}{#2}{#3}}


\def\to{\mathop{\rightarrow}}


\def\myint{\int\mkern-5mu}
\def\frac#1#2{{{#1} \over {#2}}\,}  
\def\sfrac#1#2{{\textstyle\frac{#1}{#2}}}  


\def\Dsl{\hbox{\kern.1em/\kern-.7000em$D$}} 



\def\twi{\widetilde}

\def\scr#1{{\cal #1}}

\def\mybar#1{\kern 0.8pt\overline{\kern -0.8pt#1\kern -0.8pt}\kern 0.8pt}
\def\sla#1{\raise.15ex\hbox{$/$}\kern-.57em #1}
\def\Sla#1{\kern.15em\raise.15ex\hbox{$/$}\kern-.72em #1}

\def\roughly#1{\mathrel{\raise.3ex\hbox{$#1$\kern-.75em%
    \lower1ex\hbox{$\sim$}}}}




\def\avg#1{\langle #1 \rangle}



\hyphenation{ano-ma-ly ano-ma-lies}
\hyphenation{ba-ry-on ba-ry-ons}
\hyphenation{la-gran-gian la-gran-gians}
\hyphenation{phy-sics phy-si-cal}

\def\al{\alpha}

\def\Del{\Delta}
\def\gam{\gamma}
\def\Gam{\Gamma}

\def\lam{\lambda}
\def\Lam{\Lambda}

\def\sig{\sigma}
\def\Sig{\Sigma}

\def\ChPT{\raise.45ex\hbox{$\chi$}PT}

\def\rhs{right-hand side}

\def\hc{{\rm h.c.}}




\id
IASSNS-HEP-94-107, LBL-36599, MIT-CTP-2395
\endid
\rightline{hep-ph/9501233}

\title
\vskip-0.05in
\centerline{New Symmetries of Supersymmetric}
\vskip.07in
\centerline{Effective Lagrangians}
\endtitle
\vskip.2in


\authors
\centerline{Markus A. Luty\footnote{$^*$}
{e-mail: luty@ctp.mit.edu}}
\vskip .1in
\centerline{\it Center for Theoretical Physics}
\centerline{\it Massachusetts Institute of Technology}
\centerline{\it Cambridge, Massachusetts 02139}
\vskip .1in
\centerline{John March--Russell\footnote{$^\dagger$}
{e-mail: jmr@guinness.ias.edu}}
\vskip .1in
\centerline{\it School of Natural Sciences}
\centerline{\it Institute for Advanced Study}
\centerline{\it Olden Lane}
\centerline{\it Princeton, New Jersey 08540}
\vskip .1in
\centerline{Hitoshi Murayama\footnote{$^\ddagger$}
{On leave of absence from Department of Physics,
Tohoku University, Sendai, 980 Japan;
e-mail: murayama@theorm.lbl.gov}}
\vskip .1in
\centerline{\it Theoretical Physics Group}
\centerline{\it Lawrence Berkeley Laboratory}
\centerline{\it 1 Cyclotron Road}
\centerline{\it Berkeley, California 94720}
\endauthors

\abstract
We consider the structure of effective lagrangians describing the
low-energy dynamics of supersymmetric theories in which a global symmetry $G$
is spontaneously broken to a subgroup $H$ while supersymmetry is unbroken.
In accordance with the supersymmetric Goldstone theorem, these lagrangians
contain Nambu--Goldstone superfields associated with a coset space
$G^{\rm c} / \hat H$, where $G^{\rm c}$ is the complexification of $G$ and
$\hat H$ is the largest subgroup of $G^{\rm c}$ that leaves the order
parameter invariant.
The lagrangian may also contain additional light matter fields.
To analyze the effective lagrangian for the matter fields, we first consider
the case where the effective lagrangian is obtained by integrating out heavy
modes at weak coupling (but including non-perturbative effects such as
instantons).
We show that the superpotential of the matter fields is $\hat{H}$ invariant,
which can give rise to non-trivial relations among independent $H$-invariants
in the superpotential.
We also show that the K\"ahler potential of the matter fields can be
restricted by a remnant of $\hat{H}$ symmetry.
These results are non-perturbative and have a simple group-theoretic
interpretation.
When we relax the weak-coupling constraint, there appear to be additional
possibilities for the action of $\hat{H}$ on the matter fields, hinting that
the constraints imposed by $\hat{H}$ may be even richer in strongly coupled
theories.
\endabstract


\vfill\eject


\def\susy{supersymmetry}
\def\susc{supersymmetric}
\def\rep{representation}
\def\leff{effective lagrangian}

\def\sp{superpotential}

\def\dt{d^2\theta\,}
\def\dtb{d^2\mybar\theta\,}
\def\dtt{\dt\dtb}

\def\Gc{G^{\rm c}}
\def\Hc{H^{\rm c}}
\def\Kc{K^{\rm c}}
\def\Hh{\hat{H}}
\def\Kh{\hat{K}}

\def\gz{g_0^{\vphantom{1}}}

\def\rep{representation}
\def\trans{transformation}

\section{Introduction}
Supersymmetry provides an elegant framework for understanding the
hierarchy between the weak scale and much larger mass scales such as the
grand-unification and Planck scales that are believed to play a fundamental
role in nature \ref\hier{
M. Veltman, {\sl Acta Phys. Pol.}\/ {\bf B12}, 437 (1981);
L. Maiani, in {\it Proceedings of the Eleventh Gif-sur-Yvette Summer
School on Particle Physics}, Gif-sur-Yvette, France, 1979 ({\sl Inst. Nat.
Phys. Nucl. Phys. Particules},\/ Paris, 1980), p.3;
S. Dimopoulos and S. Raby {\sl Nucl. Phys.}\/ {\bf B192}, 353 (1981);
E. Witten, {\sl Nucl. Phys.}\/ {\bf B188}, 513 (1981);
M. Dine, W. Fischler and M. Srednicki, {\sl Nucl. Phys.}\/ {\bf B189}, 575
(1981).}.
However, there is at present no direct information about what role
\susy\ will play in the more fundamental theory that we believe lies behind
the standard model of electroweak and strong interactions.
Given this situation, we believe it is essential to understand the general
features of \susc\ theories as fully as possible.

In many models for physics beyond the standard model, the symmetries
(approximate, exact, or gauged) that we observe are remnants of a larger
symmetry that is spontaneously broken at some high energy scale $\Lam$.
Because the scale $\Lam$ is often too large to be probed directly, it is
important to know what constraints this places on the physics at observable
energies $E \ll \Lam$.
For non-\susc\ theories, this question was answered in an elegant paper by
Coleman, Wess, and Zumino
\ref\CWZ{S. Coleman, J. Wess, and B. Zumino, \PRV{117}{2239}{1969}.}.
This paper derives a useful canonical form for the most general \leff\
describing the low-energy physics in a model where a global symmetry $G$ is
broken spontaneously down to a subgroup $H$.
The \leff\ contains fields for the Nambu--Goldstone bosons (NGB's)
associated with the coset space $G/H$, as well as additional light ``matter''
fields that can be chosen to transform according to linear \rep s of $H$.
The matter fields can couple to each other in the most general way allowed by
$H$ invariance, while the NGB's are derivatively coupled
\ref\CCWZ{C. Callan, S. Coleman, J. Wess, and B. Zumino,
\PRV{117}{2239}{1969}.},
so their interactions are suppressed by powers of $E / \Lam$.
Therefore, at sufficiently low energies, the only important interactions are
those of the matter fields among themselves.
Since the matter fields can interact in the most general way allowed by
the unbroken group $H$, one can describe this result by saying that $H$
invariance is the only remnant of the symmetry group $G$ at energy scales
small compared to $\Lam$.

In this paper, we show that this result is modified in an interesting way
in \susc\ theories.
We consider a theory with $N = 1$ \susy\ in which a symmetry group $G$ is
spontaneously broken down to a subgroup $H$, while \susy\ is left
unbroken.\footnote{$^*$}
{We do not treat the case where a $U(1)_R$ symmetry is spontaneously broken.}
We consider the most general \leff\ describing the interactions of NGB's and
their superpartners (which we collectively refer to as SNGB's), and ``matter''
chiral superfields.
The SNGB's are described by chiral superfields living in the coset space
$\Gc / \Hh$, where $\Gc$ is the complexification of $G$ and $\Hh$ is the
largest subgroup of $\Gc$ that leaves the order parameter invariant;
this is in agreement with the \susc\ Goldstone theorem
\ref\susygold{W. Lerche, \NPB{248}{475}{1984}.}.
Clearly, $\Hh \supseteq \Hc$, but $\Hh$ is in general {\it larger} than $\Hc$
\ref\Lfields{W. Lerche, \NPB{238}{582}{1984}.}\ref\BKMU{M. Bando,
T. Kuramoto, T. Maskawa, and S. Uehara,
\PTP{72}{313}{1984}.}. (The special case where $\Hh = \Hc$ was discussed
extensively in the literature; see {\it e.g.}\/, \ref\doubling{W.
Buchmuller, R. Peccei and T. Yanagida, \NPB{227}{503}{1983}; G. Shore,
\NPB{248}{123}{1984}; W. Lerche and D. Lust, \NPB{244}{157}{1984}.}.)

To analyze the matter fields, we begin by discussing the case where the \leff\
is obtained by integrating out heavy modes at weak coupling.
Our results rely only on symmetry arguments, and are therefore valid
non-perturbatively.
This is important despite the fact that non-perturbative effects in weak
coupling vanish faster than any power of the coupling (instanton effects, for
example).
This is because non-perturbative effects in \susc\ theories can lift
degeneracies that persist to all orders in perturbation theory
\ref\pertnr{M.~T.~Grisaru, W.~Siegel and M.~Rocek,
\NPB{159}{429}{1979}.}
\ref\Witten{E. Witten, \NPB{185}{513}{1981}.}.
Many of the non-perturbative effects in \susc\ gauge theories discussed in
the recent literature
\ref\Splb{N. Seiberg, \PLB{318}{469}{1993}.}\ref\Sgauge{See for example:
N. Seiberg, \PRD{49}{6857}{1994};
K. Intriligator, R. G. Leigh, and N. Seiberg, \PRD{50}{1092}{1994}.}\
(see also \ref\ADS{I. Affleck, M. Dine, and N. Seiberg \NPB{241}{493}{1984}})
are interesting examples of this phenomenon.

For the weak-coupling case, the matter fields transform according to linear
\rep s of the group $\Hh$, even though the true unbroken symmetry of the
theory is $H$.
Supersymmetry restricts the way that $\Hh$ is broken down to $H$, and our
first major result is that holomorphy implies that the effective \sp\ of the
matter fields is in  fact $\Hh$ invariant.
Because $\Hh$ can be larger than $\Hc$, this can lead to non-trivial relations
between different $H$-invariants in the effective \sp.
Perhaps more surprisingly, we show that there can be a remnant of $\Hh$
symmetry that restricts the effective K\"ahler potential of the matter fields
as well.
We illustrate these results with simple explicit models.

\ignore
This result implies that the statement of ``strong naturalness'' is modified in
\susc\ theories, in the following sense:
If we start with a theory with a symmetry $G$ and all couplings allowed by
the symmetry are order 1 (in some ``natural'' units), then if $G$ is broken
down to $H$, then it may happen that the low-energy theory has has relations
among different $H$-invariants.
When we write a model (at the weak scale, for example) with a symmetry $H$,
we may wish to remain agnostic about the question of whether $H$ may be the
unbroken remnant of a larger symmetry $G$ spontaneously broken at a higher
(experimentally inaccessible) scale.
In that case, imposing the kinds of relations among $H$ invariants discussed
in this paper would appear to be completely natural, since the relations follow
entirely from the pattern of symmetries and their breaking.
\endignore

We then relax the assumption of weak coupling in the fundamental theory and
consider the most general effective lagrangian describing the low-energy
dynamics when $G$ is spontaneously broken to $H$.
We are unable to classify the group action of $\Hh$ on the effective fields
in this case:
for example, there are cases where the $\Hh$ action cannot be made linear
by redefining the effective fields.
Even if $\Hh$ acts linearly, there are $\Hh$ \rep s for which we are
unable to write kinetic terms.
While it is certainly dangerous to draw any conclusions from ignorance, we
note that this may be taken as a hint that the role of $\Hh$ may be even
richer in strong-coupling theories.

This paper is organized as follows:
in section 2, we consider the most general \leff\ that can describe the
low-energy dynamics of the spontaneously broken theory.
We explain the role of the groups $\Gc$ and $\Hh$ and give some results on
the structure of these groups.
In section 3, we turn our attention to the matter fields and derive a
simple canonical form for the \leff\ describing the SNGB's and matter fields
for the case where the effective theory is obtained by integrating out
heavy modes at weak coupling;
this section contains the main results of this paper.
In section 4, we analyze the most general effective lagrangian describing
spontaneous symmetry breaking.
Section 5 contains our conclusions.

\section{The effective lagrangian for the SNGB's}
In this section, we consider the most general \leff\ describing
the low-energy dynamics of a theory with a compact global symmetry group $G$
spontaneously broken to a (compact) subgroup $H$, while \susy\ is left
unbroken.
We will concentrate on the SNGB sector of the \leff\ in this section, leaving
a detailed discussion of the matter fields for the next two sections.

\subsection{The Role of $\Gc$ and $\Hh$}
The main new feature of the \susc\ case is that the group $\Gc$ plays an
important role in restricting the low-energy couplings.
$\Gc$ is the complexification of $G$, defined by choosing a hermitian basis of
generators for $G$ and allowing the group parameters to be complex.
To understand the importance of this group, consider the underlying
``fundamental'' theory whose dynamics gives rise to the symmetry breaking.
We assume that this theory is a $N = 1$ \susc\ theory of chiral superfields
coupled to gauge superfields.
We can write the lagrangian for this theory as
\eq
\label\lfund
\scr L_{\rm fund} = \myint \dtt K(\mybar\Phi, \Phi)
+ \left( \myint \dt W(\Phi) + \hc \right),
\eeq
where we have shown only the dependence on the chiral superfields $\Phi$;
gauge fields are also present in general, but are not written explicitly.
This lagrangian is assumed to have a global symmetry $G$, which must of
course commute with the gauge group.

The first observation is that the superpotential $W(\Phi)$ is actually
invariant under $\Gc$
\ref\OWess{B. Ovrut and J. Wess, \PRD{25}{409}{1982}.}.
The reason is simply that $W$ is a holomorphic function of $\Phi$ (that is,
it is independent of $\bar\Phi$), and so it is invariant whether the group
parameters are taken real or complex.

The K\"ahler potential $K(\Phi, \bar\Phi)$ is not holomorphic, and is
therefore not invariant under $\Gc$.
However, we can make the K\"ahler potential formally invariant under $\Gc$ by
introducing ``spurion'' gauge field sources $\scr V$ transforming under $\Gc$
as
\eq
\label\vtrans
e^{\scr V} \mapsto g^{-1\dagger} e^{\scr V} g^{-1},
\qquad g \in \Gc.
\eeq
These gauge fields are not dynamical, and we will set $\scr V = 0$ at the
end of the calculation.\footnote{$^*$}
{This is analogous to the treatment of anomalies by Wess and
Zumino \ref\WZ{J. Wess and B. Zumino, \PLB{37}{95}{1971}.}.
In this case, global symmetries are enlarged to gauge symmetries
by introducing spurion gauge fields, and it is required that the
low-energy effective lagrangian has the same anomalous properties as
the microscopic lagrangian.
Even when one sets all the spurion gauge fields to zero, one is still left
with a non-trivial Wess--Zumino term.
In this paper we will not address the issue of the appearance of such terms
in the \susc\ \leff\
\ref\NemRohm{See for example: D. Nemeschansky and R. Rohm,
\NPB{249}{157}{1985}.}.}
(Differentiating with respect to components of $\scr V$ allows us to obtain
information about symmetry currents and related operators, and is also useful
for making contact with the ``current algebra'' approach to the low-energy
dynamics.)
We can then write the formally $\Gc$-invariant lagrangian by replacing
$\mybar\Phi$ with
\eq
\mybar\Phi e^{\scr V} \mapsto \mybar\Phi e^{\scr V} \cdot g^{-1},
\qquad g \in \Gc.
\eeq
(If there are derivatives in $K$, they must be replaced by gauge-covariant
derivatives constructed from $\scr V$.)
The role of $\scr V$ is to keep track of how $\Gc$ is explicitly broken down
to $G$ by the K\"ahler terms in the fundamental
lagrangian.

To understand why this is a useful thing to do, it is helpful
to contrast our introduction of $e^{\scr V}$ with the more familiar
case of explicit flavor symmetry breaking by current quark masses in QCD.
In QCD with $N_F$ quark flavors there is a $SU(N_F)_L \times SU(N_F)_R$
chiral symmetry that is explicitly broken by quark masses.
The effects of this explicit breaking are taken into account by treating the
quark mass $m_q$ as a spurion field transforming under
$SU(N_F)_L \times SU(N_F)_R$ as
\eq
m_q \mapsto L m_q R^\dagger,
\eeq
which formally restores the chiral symmetry of the QCD lagrangian.
This is useful if the quark masses are small (compared to
$\Lambda_{\rm QCD}$), because terms proportional to many powers of $m_q$
in the low-energy effective lagrangian below the scale $\Lam_{\rm QCD}$
can then be neglected.
If the quark masses are not small, introducing the quark mass as a spurion is
not useful, since many powers of $m_q$ can be used to write down
any desired $SU(3)$-violating term with an unsuppressed coefficient.

In the \susc\ case, the symmetry $\Gc$ is not an approximate symmetry because
$e^{\scr V}$ is not small in any sense.
Nevertheless, it is useful to introduce the gauge field spurion explicitly
because one cannot use it to write down arbitrary $G$-invariant terms in the
\leff.
To see this, note that only terms with no (spacetime or \susy) derivatives
acting on $\scr V$ are non-zero when we set $\scr V = 0$, so we can
restrict attention to such terms.
But {\it functions of $\scr V$ that do not involve derivatives of $\scr V$
cannot appear in the effective superpotential\/}, because their \trans\
properties involve $g^\dagger$, which is an antichiral superfield.
Therefore, {\it the superpotential of the \leff\ behaves as though the
underlying theory were invariant under} $\Gc$.
Furthermore we will argue in subsection~3.4 that the dependence of
the K\"ahler potential on $\scr V$ is
restricted by the spurious gauge transformation
properties of $e^{\scr V}$, and we find that {\it the K\"ahler
potential of the \leff\ is also restricted by a remnant of
$\Gc$ symmetry\/}. These are the general principles behind our results;
we will see them in action repeatedly below.

The group $\Hh$ is defined to be the largest unbroken subgroup of $\Gc$.
To be precise, we assume that $G$ is spontaneously broken by an order
parameter $v$ that can be thought of as an element of a (reducible) \rep\
$\rho$ of $G$.
We can extend $\rho$ to a \rep\ of $\Gc$ simply by allowing the group
parameters of $G$ to be complex.
The \rep\ matrices therefore do not depend on the complex conjugates of the
group parameters, so $\rho$ can be thought of as a holomorphic \rep\ of $\Gc$.
The group $\Hh$ is then defined by
\eq
\label\Hhdef
\Hh \equiv \{ g \in \Gc \;|\; \rho(g) v = v \}.
\eeq
That is, $\Hh$ can be viewed as the unbroken subgroup of $\Gc$;
of course, $\Hh$ is broken explicitly down to $H$ by the spurion gauge
field $e^{\scr V}$.
We note that $\Hh \supseteq \Hc$, but we will see that $\Hh$ is in general
{\it larger} than $\Hc$ \BKMU.
We will describe the structure of $\Hh$ in more detail in subsection~2.3.

\subsection{The Effective Lagrangian}
We now turn to the general structure of the low-energy \leff.
We begin by discussing the conditions on the \leff\ that encode the fact that
it describes the low-energy dynamics of a theory where a global symmetry $G$
is spontaneously broken down to a subgroup $H$, while \susy\ is unbroken.
First, the \leff\ must be \susc, so we assume that it can be written in
terms of chiral superfields. (Light gauge superfields can be
introduced by gauging part or all of the global $G$ symmetry.
This will not be discussed here.)

Second, since the original theory (including the field $\scr V$) is invariant
under $\Gc$, there is a $\Gc$ action on the fields of the \leff\ that is
nonlinear in general, and which we write as
\eq
\Phi \mapsto T(g)(\Phi),
\qquad g \in \Gc
\eeq
with
\eq
T(g_1 g_2)(\Phi) = T(g_1)(T(g_2)(\Phi)), \qquad
T(1)(\Phi) = \Phi.
\eeq
We assume that the \leff\ is invariant under this \trans.
The effective theory also contains the spurion gauge field $\scr V$
transforming as in eq.~\vtrans, which breaks $\Gc$ explicitly down to $G$.

Finally, we must also encode the information that the symmetry $G$ is broken
spontaneously by the order parameter $v$ (introduced above).
We want to interpret the fields in the \leff\ as fluctuations about the vacuum
described by the order parameter $v$, so we demand that the target space
(space of fields) in the \leff\ contain a special point (the origin) that is
preserved by the action of the subgroup $\hat{H}$.
Here we are implicitly assuming that the complex structure of the full theory
is inherited by the effective theory, that is, that there are no ``holomorphic
anomalies'' in the matching that determines the \leff.
Since this matching is infrared safe, this is a reasonable assumption
\ref\SV{M.~A.~Shifman and A.~I.~Vainshtein, \NPB{277}{456}{1986}.}.

We now consider the most general \leff\ satisfying the assumptions above.
We will follow closely the arguments of ref.~\CWZ.
The basic idea is to use the freedom to make field redefinitions to put the
\leff\ into a canonical form where its physical content is manifest.
Specifically, if we make a field redefinition of the form
\eq
\Psi = \Phi F(\Phi),
\eeq
with $F(0) = 1$ (that is, the redefinition preserves the origin of field
space), then the physics described by the the \leff s written in terms of
$\Phi$ and $\Psi$ is identical.

We therefore make such a field redefinition by decomposing the target space
into the orbits of the origin under $\Gc$ and the rest.
Specifically, we write
\eq
\Phi = T(\xi)(\Psi),
\eeq
where
\eq
\xi = e^{i\Pi} \in \Gc / \Hh,
\eeq
and $\Psi$ are coordinates for the part of the target space that is left
invariant under $\hat{H}$.
We can see how the new fields $\xi$ and $\Psi$ transform by noting that for
any $g \in \Gc$
\eq
\Phi \mapsto T(g)(T(\xi)(\Psi)) = T(g\xi)(\Psi).
\eeq
We then decompose
\eq
g \xi = \xi'(g, \xi) \hat{h}(g, \xi),
\qquad \xi' \in \Gc / \Hh,\ \ \ \hat{h} \in \Hh
\eeq
and write
\eq
\Phi \mapsto T(g \xi \hat{h}^{-1}(g, \xi))(T(\hat{h}(g, \xi)(\Psi))).
\eeq
That is, the fields $\xi$ and $\Psi$ transform as
\eqa
\label\xitrans
\xi &\mapsto g \xi \hat{h}^{-1}(g, \xi), \eol
\label\Psitrans
\Psi &\mapsto T(\hat{h}(g, \xi))(\Psi). \eeol
\eeq
The \leff\ also contains the spurion gauge field
transforming as in eq.~\vtrans.

We see that with our assumptions, the \leff\ automatically contains fields
$\xi$ that live in the coset space $\Gc / \Hh$.
One can check that the fields $\xi$ couple to broken symmetry currents in the
manner required by the \susc\ version of Goldstone's theorem \susygold
(see also \Lfields \doubling), so that
we can identify them with the SNGB's.
The fields $\Psi$ are identified with light ``matter'' fields.

The existence of the fields $\xi \in \Gc / \Hh$ is a direct consequence of our
ability to formally promote $\Gc$ to a symmetry of the fundamental lagrangian
by introducing the gauge spurion $\scr V$.
Therefore, as a consistency check, we should understand why the presence of
$\scr V$ in the \leff\ does not allow us to write a mass term for the SNGB's
in a theory with no matter fields.
(For example, a quark mass spurion in QCD allows us to write mass terms for
the NGB's.)
A mass term for the SNGB's must be a superpotential term with no derivatives
(spacetime or \susy).
It is easy to see that the constraints of the \trans\ rules in eqs.~\vtrans\
and \xitrans, together with the requirement of holomorphy, imply that no
such term is possible.

We will not discuss the structure of the effective
lagrangian for the SNGB's in much detail, but we briefly indicate
how to write an invariant kinetic term for
the SNGB's.
We restrict ourselves to groups $\Hh$ for which
\eq
\rho(\hat{h}^\dagger) = \rho(\hat{h})^\dagger.
\eeq
We will see in subsection~3.4 how this condition can fail,
and how to generalize the construction below to all $\Hh$.
We then define
\eq
e^{\scr W} \equiv \xi^\dagger e^{\scr V} \xi \in \Gc,
\eeq
which transforms like a gauge field for the group $\Hh$:
\eq
e^{\scr W} \mapsto \hat{h}^{-1\dagger}(g, \xi) e^{\scr W}
\hat{h}^{-1}(g, \xi).
\eeq
Then we can write the kinetic term
\eq
\label\kterm
\scr L_{\rm eff} = \myint \dtt v^\dagger \rho(e^{{\scr W}}) v,
\eeq
where $v$ is the order parameter in the \rep\ $\rho$ of $\Gc$ (see eq.~\Hhdef).
To see that eq.~\kterm\ contains a kinetic term for the SNGB's, note that
$\rho(\xi) v = i\rho(\Pi) v + O(\Pi^2)$, so that
\eq
\scr L_{\rm eff} = \myint \dtt | \rho(\Pi) v|^2 + O(\scr V) + O(\Pi^3).
\eeq
Note that $\rho(\Pi) v$ is linear in $\Pi$ and is nonzero for all $\Pi \ne 0$
by the definition of the SNGB fields, completing the
argument.\footnote{$^*$}{Similar kinetic terms to eq.~\kterm\ were discussed
in refs.~\Lfields\BKMU. Note that eq.~\kterm\ is not of the form
proposed in ref.~\ref\Zumino{B.~Zumino, \PLB{87}{203}{1979}.}. This
form of the K\"ahler potential can never appear in a consistent \leff\
describing the dynamics of SNGB fields \ref\BL{W. Buchmuller and
W. Lerche, {\sl Ann. Phys.} {\bf 175}, 159 (1987)}
\ref\KS{A. Kotcheff and G. Shore,
{\sl Int. J. Mod. Phys.} {\bf A4}, 4391 (1989).}.}
We could go on to discuss the general form of the K\"ahler potential for the
SNGB's and the resulting low-energy theorems, but the main focus of this paper
is on the matter fields, so we will leave these topics for the future.

\subsection{Structure of $\Hh$}
We now give some results on the structure of the group $\Hh$, and illustrate
them with some simple examples.
The main structure theorem is that $\Hh$ has the Levy decomposition
\eq
\label\Levy
\Hh = \Kc \wedge N,
\eeq
where ``$\wedge$'' denotes a semidirect product (with $\Kc$ acting on $N$).
Specifically, this means that any $\hat h \in \Hh$ can be uniquely decomposed
as $\hat h = k n$ with $k \in \Kc$, $n \in N$, and that $k n k^{-1} \in N$ for
any $k \in \Kc$ and $n \in N$.
Here, $K$ is a compact group that can be written as a direct product of a
semisimple group and an abelian group, and $N$ is a unipotent group:
that is, $N$ is isomorphic to a group of upper-triangular matrices with
$1$'s on the diagonal.
(This is the ``algebraic'' version of the Levy decomposition;
see \eg\ ref.~\ref\Borel{A. Borel, {\it Linear Algebraic Groups}, (Springer,
1991).}.
In the context of SNGB's, this result is discussed in ref.~\BKMU.)
The multiplication law for the Levy factors of $\Hh$ is (in obvious notation)
\eq
\label\Hhmult
\hat{h}_1 \hat{h}_2 = k_1 n_1 \cdot k_2 n_2
= (k_1 k_2) \cdot (k_2^{-1} n_1 k_2 n_2)
\eeq
where $k_1 k_2 \in \Kc$ and $k_2^{-1} n_1 k_2 n_2 \in N$.


As the examples below will make clear, $H \subseteq K$, but $K$ can be larger
than $H$.
(In fact, $K \not\subset G$ in general.)
Thus, $\Hh \supset H$ if either $N \ne 0$ or $K \supset H$.
We illustrate both of these possibilities below.

Consider first an example with $G = U(N)$ broken by an order parameter
$\avg\Phi$ in the defining \rep\ of $U(N)$.
We can make a $U(N)$ \trans\ to put $\avg{\Phi}$ in the standard form
\eq
\label\exvev
\avg\Phi = \pmatrix{v \cr 0 \cr \vdots \cr 0 \cr},
\eeq
and it is clear that the unbroken group is $H = U(N - 1)$.
The group $\Hh$ is given by the set of all $N \times N$ matrices of the block
form
\eq
\label\exHh
\hat{h} = \bordermatrix{   & 1 & N - 1 \cr
                 \ \ \ \ 1 & 1 &   a   \cr
                     N - 1 & 0 &   u   \cr},
\eeq
where $u \in U(N - 1)^{\rm c} = GL(N - 1, C)$ and $a$ is a general complex
row vector with $N - 1$ entries.
The entries in $a$ are allowed to be non-zero because elements of $\Hh$ are not
required to be unitary (equivalently, the generators of $\Hh$ are not
required to be hermitian).
Therefore, in this example $K = H = U(N - 1)$, and $N$ is the group of
matrices of the form
\eq
n = \bordermatrix{   & 1 & N - 1 \cr
           \ \ \ \ 1 & 1 &   a   \cr
               N - 1 & 0 &   1   \cr}.
\eeq
According to the arguments given earlier in this section, there is one real
scalar SNGB for each generator of $\Gc / \Hh$.
It is easy see that
\eq
{\rm dim}\, G / H = 2N - 1, \qquad
{\rm dim}\, \Gc / \Hh = 2N.
\eeq
Therefore, there is one ``extra'' SNGB whose superpartner is a NGB.
When we discuss explicit models in subsection 3.5, we will see that the
SNGB's that are not NGB's can be identified with excitations along flat
directions of the potential that gives rise to the spontaneous symmetry
breaking.

Next, consider an example with $G = U(N)$ as before, but with two order
parameters (or equivalently, an order parameter in a reducible \rep\ of $G$).
The order parameters are $\avg{\Phi_+}$ in the defining \rep,
and $\avg{\Phi_-}$ in the complex conjugate of the defining \rep.
We can use $U(N)$ \trans s to put the order parameters in the standard form
\eq
\avg{\Phi_+} = \pmatrix{v_+ \cr 0 \cr 0 \cr \vdots \cr 0 \cr}, \qquad
\avg{\Phi_-} = \pmatrix{v_- \cr w \cr 0 \cr \vdots \cr 0 \cr},
\eeq
and it is clear that the unbroken group is $H = U(N - 2)$.
To see what $\Hh$ is in this case, note that we can use a
$U(N)^{\rm c} = GL(N, C)$ \trans\ to further simplify the order parameters:
if we choose
\eq
\label\gzex
g_0 = \bordermatrix{   & 1 & N - 1 \cr
             \ \ \ \ 1 & 1 &   a   \cr
                 N - 1 & 0 &   1   \cr} \in U(N)^{\rm c}, \qquad
a^T = \pmatrix{ w / v_- \cr 0 \cr \vdots \cr 0 \cr},
\eeq
then
\eq
\label\gzvevex
g_0 \avg{\Phi_+} = \pmatrix{v_+ \cr 0 \cr \vdots \cr 0 \cr}, \qquad
g_0^{-1 T}\avg{\Phi_-} = \pmatrix{v_- \cr 0 \cr \vdots \cr 0 \cr}.
\eeq
This shows that $\Hh$ is given by all matrices of the form
\eq
\label\Hhex
\hat{h} = g_0^{-1} \hat{k} \gz, \qquad
\hat{k} = \bordermatrix{   & 1 & N - 1 \cr
                 \ \ \ \ 1 & 1 &   0   \cr
                     N - 1 & 0 &   u   \cr},
\eeq
where $u \in U(N - 1)^{\rm c}$.
Therefore, in this example $N = 0$ and $K \simeq U(N - 1)$.
Note that $K \not\subset G$ in this case (although $K$ is {\it isomorphic}
to a subgroup of $G$).
It is easy see that
\eq
{\rm dim}\, G / H = 4N - 4, \qquad
{\rm dim}\, \Gc / \Hh = 2(2N - 1).
\eeq
Therefore, there are 2 ``extra'' SNGB's in this case.
As in the previous example, when we discuss explicit models, we will see that
they can be identified with excitations along flat directions of the potential.

We will see in subsection 3.5 that these symmetry breaking patterns can arise
in simple toy models, and that they give rise to interesting restrictions on
the low-energy \leff.

\section{Matter Fields: Weak Coupling}
In this section, we consider the \leff\ including the matter fields in the
case where the \leff\ is obtained by integrating out heavy modes at weak
coupling.
The reason for making this restriction is that in this case, the group
$\Hh$ acts linearly on the matter fields in the \leff, and we
will be able to obtain a simple canonical form for the \leff:
we find that the superpotential for the matter fields is invariant under $\Hh$,
while the K\"ahler potential is constrained by $\Kc$, both of which are in
general larger than $\Hc$.
(As we will see in the next section, the general case is more complicated.)

\subsection{Transformation of the Effective Fields}
We consider a ``fundamental'' theory with chiral superfields $\Phi$ invariant
under a global symmetry $G$.
Because the theory is weakly coupled, the order parameter can be taken to be
$\avg\Phi$.
We can therefore write
\eq
\label\effields
\Phi = \rho(\xi) \left[ \avg\Phi + \Psi + \Del \right],
\eeq
where $\rho$ is the \rep\ (reducible in general) of $G$ under which $\Phi$
transforms, $\xi \in \Gc / \hat{H}$ parameterizes the excitations of $\Phi$ in
the broken symmetry directions (in the generalized sense of $\Gc$ invariance),
and $\Psi$ and $\Del$ are the excitations of $\Phi$ in the remaining
directions.
We assume that the fields $\Psi$ remain light (relative to
$\avg\Phi$), while the fields $\Del$ get masses of order
$\avg\Phi$. (For simplicity, we assume that there is a single scale
set by $\avg\Phi$. The extension to the case where $\avg\Phi$ contains
several different scales should be clear.)
We then imagine computing an effective action containing only the light
degrees of freedom by integrating out the heavy fields $\Del$.
The fields in the resulting low-energy \leff\ will transform under $\Gc$ as
\eqa
\label\xitrans
\xi &\mapsto g \xi \hat{h}^{-1}(g, \xi), \eol
\label\mtrans
\Psi & \mapsto R(\hat{h}(g, \xi)) \Psi, \eeol
\eeq
where $R$ is the $\Hh$ \rep\ obtained by reducing the \rep\ $\rho$ of $\Gc$.
The \leff\ can be constructed by writing down the most general form allowed by
the symmetries and then determining the coefficients by matching onto the
fundamental theory.
We address the first part of the problem in this section, concentrating on the
matter fields $\Psi$.

The striking fact about eq.~\mtrans\ is that the matter fields transform
according to \rep s of $\Hh$, even though the true unbroken symmetry of the
theory is only $H$.
As discussed in subsection 2.1, the reason for this is the fact that \susy\
constrains the way the field $\scr V$ breaks $\Hh \to H$.

\subsection{The Effective Superpotential}
As pointed out in subsection 2.1, the \trans\ rule for $\scr V$ in
eq.~\vtrans\ does not allow $\scr V$ to appear in the effective superpotential
unless derivatives act on $\scr V$.
Since such terms vanish when $\scr V = 0$, the effective superpotential is
invariant under $\Hh$.
In cases where $\Hh$ is larger than $\Hc$, this leads to additional
restrictions on the effective superpotential beyond those imposed by $\Hc$
invariance.

As a simple example, consider a theory with the symmetry breaking pattern of
the first example in subsection 2.3 (see the discussion surrounding
eqs.~\exvev\ and \exHh).
That is, $G = U(N)$, $H = U(N - 1)$, and the order parameter is in the
defining \rep\ of $U(N)$.
Suppose now that the low-energy theory contains matter fields $\Psi_+$
transforming according to the defining \rep\ of $U(N)$.
Under $U(N - 1)$, $\Psi_+$ decomposes into a singlet and a defining \rep,
but $\Hh$ invariance mixes these \rep s, leading to restrictions on the
effective superpotential.
For example, if the low-energy theory also contains matter fields $\Psi_-$
transforming according to the complex conjugate of the defining \rep\ of
$U(N)$, then we can write
\eq
\Psi_\pm = \bordermatrix{ & \cr \ \ \ \ 1 & A_\pm \cr N - 1 & B_\pm \cr},
\eeq
and the most general $\Hh$-invariant quartic terms in the effective
superpotential are
\eq
(\Psi_+ \Psi_-)^2 = (A_+ A_-)^2 + 2 (A_+ A_-) (B_+ B_-) + (B_+ B_-)^2, \qquad
\Psi_+ \Psi_- A_-^2, \qquad
A_-^4.
\eeq
In the first term, the relative coefficients of three $H$ invariants are fixed
by $\Hh$ invariance.
(Note that we cannot put the quartic term in this form by rescaling the
fields.)
Also note that the terms involving $A_+$ by itself are forbidden by $\Hh$,
even though they are allowed by $H$.
We will consider an explicit model with this structure after we have
discussed the effective K\"ahler potential.

\subsection{Structure of $\Hh$ Representations}
In order to understand the structure of the effective K\"ahler potential, we
need some general results about the $\Hh$ \rep s $R$ of the matter fields.
In the class of theories we are considering, the $\Hh$ \rep\ $R$ is
obtained by reducing a $\Gc$ \rep.
To make this precise, we write (in the sense of eq.~\effields)
\eq
\Psi = P_\Psi [\Phi - \avg\Phi],
\eeq
where $P_\Psi$ is a projection operator.
That is, we think of $\Psi$ as an element of the \rep\ space of the $\Gc$
\rep\ $\rho$ that is nonzero only in a subspace.
Because the fields $\Psi$ transform among themselves under $\Hh$, the relation
between $\rho$ and the \rep\ $R$ of $\Hh$ is
\eq
\label\Hhred
\rho(\hat h) P_\Psi = R(\hat h)
\eeq
for all $\hat h \in \Hh$.
Here, we view $R$ as acting on the state space of $\rho$, but $R$ is non-zero
only on the $\Psi$ subspace.\footnote{$^*$}
{Strictly speaking, $R$ defined in this way is not a \rep, since it is not
invertible as a linear operator on the state space of $\rho$.
However, if we define the inverse of $R$ on the subspace on which it acts
non-trivially, we can think of $R$ as a \rep.}

In the appendix, we show that any $\Hh$ \rep\ $R$ obtained by reducing a $\Gc$
\rep\ as in eq.~\Hhred\ is equivalent to a \rep\ by matrices of the block form
\eq
\label\Hhrepform
R(k n) = \pmatrix{R_1(k) & * & \cdots & * \cr
0 & R_2(k) & * & \vdots \cr
\vdots & 0 & \ddots & * \cr
0 & \cdots & 0 & R_r(k) \cr},
\eeq
where $k \in \Kc$ and $n \in N$ are the factors in the Levy decomposition of
$\Hh$.
That is, $R(\hat h)$ is an upper-block-triangular matrix with \rep\ matrices
of $\Kc$ in the diagonal blocks.
As explained in the appendix, this result can be thought of as a generalization
of Engel's theorem for the \rep s of Lie algebras.
To check that eq.~\Hhrepform\ defines a \rep\ of $\Hh$, we must use the
multiplication law for the Levy factors given in eq.~\Hhmult.
As a special case of eq.~\Hhrepform, we note that any \rep\ $R_K$ of $\Kc$
gives a \rep\ of $\Hh$, defined by $R(k n) = R_K(k)$.

\subsection{The Effective K\"ahler Potential}
We now discuss how $\scr V$ breaks $\Hh$ down to $H$ in the effective
K\"ahler potential.
Our main result is that the allowed term in the K\"ahler potential for the
matter fields are classified by $K$ invariants (not $H$ invariants).
The best way to see this is to work in a ``gauge'' for $\scr V$ where the
structure of the unbroken group is as simple as possible.
In this language, the explicit breaking of $K$ down to $H$ is accomplished by
the vacuum value of $\scr V$.

To make this precise, recall that the group $K$ defined in the Levy
decomposition eq.~\Levy\ can be larger than $H$ when there is a $\Gc$ \trans\
$g_0$ that can simplify the order parameter.
We therefore define the group
\eq
\Kh \equiv \{ \gz \hat h g_0^{-1} \;|\; \hat h \in \Hh \}.
\eeq
The group $\Kh$ is {\it isomorphic} to $\Hh$, but it is a different group of
matrices.
Maintaining this distinction is important for understanding the construction
given below.

The reason for introducing the group $\Kh$ is as follows:
if $K$ is larger than $H$, then when we choose a basis where $R$ has the form
given in eq.~\Hhrepform, we find that
$R(\hat{h}^\dagger) \ne R(\hat{h})^\dagger$.
(Here, we use the definition of $\dagger$ on $\Gc$ that makes $G$ and its
\rep s real.
The subgroup $\Hh$ and its \rep s $R$ satisfying eq.~\Hhred\ then naturally
inherit a definition of $\dagger$.)
To see why this is so, consider the second example in subsection 2.3.
The field $\Phi_+$ transforms in the defining \rep\ of $\Hh$, but this does
not have the form of eq.~\Hhrepform.
However, the defining \rep\ of $\Hh$ is equivalent to the \rep
\eq
R(\hat h) = \gz \hat h g_0^{-1} = \pmatrix{1 & 0 \cr 0 & u \cr}.
\eeq
(See eq.~\Hhex.)
This example shows that we can find a $g_0 \in \Gc$ such that
\eq
\label\Rdagg
R(g_0^{-1} \hat{h}^\dagger \gz) = R(g_0^{-1} \hat h \gz)^\dagger.
\eeq
The reason for this is that $K$ is larger than $H$ only if there is a \trans\
$g_0 \in \Gc$ that can be used to simplify the order parameter further than
can be done with $G$ \trans s alone.

When $K$ is larger than $H$, it is convenient to work with fields where
hermitian conjugation acts in a simple way.
We therefore define fields
\eqa
\twi\xi &\equiv \gz \xi g_0^{-1} \in \Gc / \Kh, \eol
\twi\Psi &\equiv \rho(g_0) \Psi, \eeol
\eeq
transforming as
\eqa
\twi\xi &\mapsto \twi g \twi\xi \hat{k}^{-1}(\twi g, \twi\xi), \eol
\twi\Psi &\mapsto \twi R(\hat{k}(\twi g, \twi\xi)) \twi\Psi, \eeol
\eeq
where
\eq
\twi g \equiv \gz g g_0^{-1} \in \Gc, \qquad
\hat{k}(\twi g, \twi\xi) \equiv \gz \hat{h}(g, \xi) g_0^{-1} \in \Kh,
\eeq
and
\eq
\twi R(\hat k) \equiv R(g_0^{-1} \hat k \gz).
\eeq
Note that with these definitions,
\eq
\label\Rpdagg
\twi R(\hat{k}^\dagger) = \twi R(\hat{k})^\dagger
\eeq
by eq.~\Rdagg.
In order to write invariants, it is useful to work in terms of the transformed
spurion gauge fields
\eqa
e^{\twi{\scr V}} &\equiv g_0^{-1\dagger} e^{\scr V} g_0^{-1}
\mapsto \twi g^{-1\dagger} e^{\twi{\scr V}} \twi g^{-1}, \eol
e^{\twi{\scr W}} &\equiv \twi\xi^{\dagger} e^{\twi{\scr V}} \twi\xi \mapsto
\hat{k}^{-1\dagger}(\twi g, \twi\xi) e^{\twi{\scr W}}
\hat{k}^{-1}(\twi g, \twi\xi).  \eeol
\eeq
These redefinitions simply amount to making a \trans\ $g_0 \in \Gc$ to
twiddled fields.
However, it is important to note that this is not a symmetry \trans, since
\eq
\label\Kbreak
e^{\twi{\scr V}} \bigr|_{\scr V = 0} = g_0^{-1\dagger} g_0^{-1}
\ne 1 {\rm\ in\ general}.
\eeq
The advantage of working in terms of these fields is that the most general
invariants are simply the most general gauge-invariant combinations of the
twiddled fields, and the explicit breaking of $\Hh$ down to $H$ is
{\it entirely} due to the vacuum value of $\twi\scr{V}$.

We can use the field $\kern.2em\twi{\phantom{\scr V}}\kern-1em\scr{W}$ to
define covariant generalizations of derivative operators, such as
\eq
\nabla_\al \equiv D_\al + e^{-\twi\scr{W}} D_\al e^{\twi\scr{W}}.
\eeq
When $\twi\scr{V}$ is replaced by its vacuum value, the derivative does not
explicitly break $\Hh$, since
\eq
e^{-\twi\scr{W}} D_\al e^{\twi\scr{W}}
= \twi\xi^{-1} e^{-\twi\scr{V}} D_\al e^{\twi\scr{V}} \twi\xi
+ \twi\xi^{-1} D_\al \twi\xi,
\eeq
and $D_\al \twi\scr{V} = 0$ when $\scr V = 0$.

We are now ready to show that the allowed terms in the effective K\"ahler
potential are classified by $K$ invariants.
This is done by constructing new matter fields $\twi\Psi_j$ that transform
according to
\eq
\twi\Psi_j \mapsto \twi R_j(k(\twi g, \twi\xi)) \twi\Psi_j,
\eeq
where $\twi R_j$ is the $j^{\rm th}$ diagonal block of the $\Kh$ \rep\
$\twi R$ (see eq.~\Hhrepform), and $k(\twi g, \twi \xi) \in \Kc$ is defined
by decomposing
\eq
\hat{k}(\twi g, \twi\xi) = k(\twi g, \twi\xi) \twi n(\twi g, \twi\xi),
\eeq
with $\twi n(\twi g, \twi\xi) \in \{ g_0 n g_0^{-1} \;|\; n \in N \}$.

We will see that the matter fields $\twi\Psi_j$ for $j = 1, \ldots, r - 1$
(where $r$ is the total number of blocks) are not holomorphic in the original
fields, so they cannot appear in the effective superpotential.
However, they can appear in the K\"ahler potential, so any $K$-invariant
function of $\twi\Psi_j$ is allowed in the K\"ahler potential.
The matter fields $\twi\Psi_j$ also depend on $\twi{\scr V}$, and when $K$ is
larger than $H$, $K$ is broken {\it only} by eq.~\Kbreak.
This will give rise to relations between the coefficients of different
$H$-invariants in the K\"ahler potential.

We begin by defining the projection operators $P_{\ge j}$ and $P_{\le j}$
acting on the \rep\ space of $\twi R$:
\eq
P_{\le j} = \bordermatrix{ & j & r - j \cr
\ \ \ \ j & 1 & 0 \cr r - j & 0 & 0 \cr}, \qquad
P_{\ge j} = \bordermatrix{ & j - 1 & r - j + 1 \cr
\ \ \ \ j - 1 & 0 & 0 \cr r - j + 1 & 0 & 1 \cr},
\eeq
where the numbers at the border of the matrix count the number of {\it blocks}
(see eq.~\Hhrepform).
\ignore
$P_{\ge j}$ is defined to project onto the subspace corresponding to all
blocks from the $j^{\rm th}$ to the last one (inclusive), while $P_{\le j}$
projects onto the subspace corresponding to all blocks from the first to the
$j^{\rm th}$ (inclusive).
\endignore
Because $\twi R(\hat k)$ is an upper-triangular matrix, it is easy to see that
\eq
P_{\ge j} \twi R(\hat k) = P_{\ge j} \twi R(\hat k) P_{\ge j}, \qquad
\twi R(\hat k) P_{\le j} = P_{\le j} \twi R(\hat k) P_{\le j},
\eeq
where we use the abbreviation $\hat k \equiv \hat{k}(\twi g, \twi\xi)$.
Similarly, $\twi R(\hat k)^\dagger$ is a lower-triangular matrix, and
hence
\eq
P_{\le j} \twi R(\hat k)^\dagger = P_{\le j} \twi R(\hat k)^\dagger
P_{\le j}, \qquad
\twi R(\hat k)^\dagger P_{\ge j} = P_{\ge j} \twi R(\hat k)^\dagger P_{\ge j}.
\eeq
We can therefore define the fields
\eq
\twi\Psi_{\ge j} \equiv P_{\ge j} \twi\Psi
\mapsto P_{\ge j} \twi{R}(\hat k) P_{\ge j} \cdot \twi\Psi_{\ge j}.
\eeq
That is, $\twi\Psi_{\ge j}$ transforms according to the \rep\
$P_{\ge j} \twi R P_{\ge j}$ consisting of the last $j$ subblocks of $\twi R$.
In particular, the matter field $\twi\Psi_r \equiv P_{\ge r}\twi\Psi$
transforms according to the \rep\ $\twi R_r$ of $K$.

We can isolate the ``middle'' blocks using the gauge field spurion.
To see how to do this, note that by eq.~\Hhred,
\eq
\twi R(e^{\twi{\scr W}}) \equiv \rho(e^{\scr W}) P_\Psi
\mapsto \twi R(\hat{k}^{-1})^\dagger \cdot \twi R(e^{\twi{\scr W}})
\cdot \twi R(\hat{k}^{-1}),
\eeq
where we have used eq.~\Rpdagg.
Thus, we can define
\eq
S_{\le j} \equiv P_{\le j} \twi R(e^{\twi{\scr W}}) P_{\le j}
\mapsto P_{\le j} \twi R(\hat{k}^{-1})^\dagger P_{\le j}
\cdot S_{\le j} \cdot
P_{\le j} \twi R(\hat{k}^{-1}) P_{\le j},
\eeq
and
\eq
S_{\ge j} \equiv P_{\ge j} \twi R(e^{-\twi{\scr W}}) P_{\ge j}
\mapsto P_{\ge j} \twi R(\hat{k}) P_{\ge j}
\cdot S_{\ge j} \cdot
P_{\ge j} \twi R(\hat{k})^\dagger P_{\ge j}.
\eeq
To see why this is useful, note that
\eq
S_{\ge j}^{-1} \twi\Psi \mapsto P_{\ge j}
\twi R(\hat{k}^{-1})^\dagger P_{\ge j}
\cdot S_{\ge j}^{-1} \twi\Psi,
\eeq
where the inverse is defined in the obvious way on the non-zero blocks.
Note that
\eq
S_{\ge j}^{-1} \twi\Psi = \twi\Psi_{\ge j} + O(\Pi) + O(\scr V),
\eeq
but $S_{\ge j}^{-1} \twi\Psi$ transforms according to a
{\it lower\/}-triangular \rep\ whose first non-zero block corresponds to
$R_j$.
Therefore, we can use the projection operator $P_{\le j}$ to construct matter
fields that transform according to $\twi R_j$:
\eq
\twi\Psi_j \equiv P^{\vphantom 1}_{\le j} S_{\ge j}^{-1} \twi\Psi
\mapsto \twi R_j(k^{-1}(\twi g, \twi\xi))^\dagger \twi\Psi_j,
\eeq
as claimed above.
Similarly, we can write invariants involving $\twi\Psi_j^\dagger$ by noting
that
\eq
\mybar{\twi\Psi}_j \equiv
\twi\Psi_j^\dagger S_{\le j}^{-1} P^{\vphantom\dagger}_{\ge j} \mapsto
\mybar{\twi\Psi}_j \twi R_j(k(\twi g, \twi\xi))^\dagger.
\eeq

Note that it is impossible to project the matter fields down further, for
example to $H$.
To see this, note that when $g_0 = 1$, $K = H$, and it is clear that we can
only project down to $K$.
When $g_0 \ne 1$, we can perform a ``gauge transformation'' to define the
``twiddled'' fields in which $g_0$ only appears in the vacuum value of the
gauge field.
However, the terms we write must respect the full $\Gc$ symmetry, and so there
are no additional invariants when the gauge fields take on particular values.

\subsection{Toy Models}
We now give some explicit toy models that illustrate the main results
obtained above, namely that the effective superpotential for the matter fields
is invariant under $\Hh$, while the effective K\"ahler potential is constrained
by $\Kc$, both of which are in general larger than $\Hc$.
In order to illustrate our results, we must consider models that spontaneously
break symmetries, and in addition contain ``matter'' fields that remain light
after symmetry breaking.
The models are therefore somewhat complicated, and we will discuss them in
two steps:
first, we construct the ``symmetry breaking sector,'' and then we add fields
to the model to get additional matter fields at low energies.

The first example has larger $K$ than $H$ and demonstrates that the
K\"ahler potential is restricted by $K$-invariance. It has $G = U(N)$
with fields
\eq
\Phi_+ \sim N, \qquad \Phi_- \sim \mybar N, \qquad \Del \sim 1.
\eeq
The most general renormalizable superpotential is
\eq
W = \frac M2 \Del^2 + \frac g3 \Del^3
+ m \Phi_+ \Phi_- + \lam \Del \Phi_+ \Phi_-,
\eeq
where we have shifted away a possible linear term in $\Del$.
It is easy to see that there are no additional accidental symmetries.
The most general vacuum of this theory is either
\eq
\avg{\Phi_\pm} = 0, \qquad
\avg\Del = 0 \ \hbox{or}\ -M / g,
\eeq
or
\eq
\avg{\Phi_+} = \pmatrix{v_+ \cr 0 \cr 0 \cr \vdots \cr 0 \cr}, \qquad
\avg{\Phi_-} = \pmatrix{v_- \cr w \cr 0 \cr \vdots \cr 0 \cr}.
\eeq
where
\eq
\label\funvac
v_+ v_- = \frac{m}{\lam^2} \left( M - \frac{gm}{\lam} \right),
\eeq
but $v_\pm$ and $w$ are otherwise arbitrary.
We will study this case.
(The flat directions are preserved to all orders
in perturbation theory \pertnr;
the techniques of ref.~\Splb\ show that this result is true even beyond
perturbation theory.) Note that we can always take $w$ to be real
and also $v_\pm$ relatively real using $G$ transformation. Therefore,
number of flat directions in this model is that corresponding to $w$ and
$v_+/v_-$, both real parameters. This precisely coincides with the
number of extra SNGB's as discussed in section 2.3.

The symmetry breaking structure is exactly the same as the second example in
subsection 2.3 (see the discussion surrounding eqs.~\gzex--\Hhex).
In this example, $\Hh \simeq U(N - 1)^{\rm c}$, so that $K = U(N - 1)$ is
larger than $H$.
There are $2N - 1$ massless chiral superfields that can all be identified
with SNGB's.
The SNGB's can be parameterized by writing
\eq
\Phi_+ = g_0^{-1} \twi{\xi} g_0 \cdot \left[ \avg{\Phi_+} + \cdots \right],
\qquad
\Phi_- = g_0^{T} \twi{\xi}^{ -1 T} g_0^{-1 T} \cdot
\left[ \avg{\Phi_-} + \cdots \right],
\eeq
where $g_0$ is defined in eq.~\gzex\ and
\eq
\twi{\xi} = e^{i\twi{\Pi}} \in \Gc / \Kh, \qquad
\twi{\Pi} = \bordermatrix{& 1 & N - 1 \cr \ \ \ \ 1 & \sig & \pi_- \cr
N - 1 & \pi_+ & 0 \cr}.
\eeq

To get a more interesting theory, we add more fields to the theory.
We write $G = SU(N) \times U(1)$ and take the fields to transform as
\eq
\eqalign{
\Phi_+ \sim (N; +1), \qquad
\Phi_- &\sim(\mybar N; -1), \qquad
\Del \sim (1; 0), \cr
\Sig_+ \sim (N; -1), \qquad
\Sig_- &\sim(\mybar N; +1), \qquad
\Gam \sim (1; -2). \cr}
\eeq
The most general dimension-4 superpotential is now
\eq
\eqalign{
W &= \frac M2 \Del^2 + \frac g3 \Del^3
+ m \Phi_+ \Phi_- + \lam \Del \Phi_+ \Phi_- \cr
&\qquad + \mu \Sig_+ \Sig_-
+ \gam \Del \Sig_+ \Sig_- + \beta \Gam \Phi_+ \Sig_-. \cr}
\eeq
There is a vacuum for which $\avg{\Phi_\pm}$ and $\avg\Del$ are as above and
\eq
\avg{\Sig_\pm} = 0, \qquad \avg\Gam = 0.
\eeq
The light fields are now the SNGB's discussed above and the matter fields
\eqa
\Psi_+ &= \bordermatrix{&\cr \ \ \ \ 1 & A_+ \cr N - 1 & B_+ \cr}
= g_0 \xi^{-1} \Sig_+, \eol
\Psi_- &= \bordermatrix{&\cr \ \ \ \ 1 & 0 \cr N - 1 & B_- \cr}
= P g_0^{-1 T} \xi^T \Sig_-, \eeol
\eeq
where
\eq
P = \bordermatrix{ & 1 & N - 1 \cr \ \ \ \ 1 & 0 & 0 \cr N - 1 & 0 & 1 \cr}.
\eeq
These fields transform as
\eq
A_+ \mapsto A_+, \qquad
B_+ \mapsto u B_+, \qquad
B_- \mapsto u^{-1 T} B_-,
\eeq
where $u \in U(N - 1)$ parameterizes
$\hat k(\twi{g}, \twi{\xi})$ as in eq.~\Hhex.
Note that $B_\pm$ reduce under the unbroken $U(N - 2)$ as a sum of a singlet
and a $(N - 2)$-dimensional \rep.
To define kinetic terms for these fields, we follow the general discussion
above and define
\eqa
S &\equiv \left[ P e^{-\twi{\scr{W}}} P \right]^{-1} \eol
&\mapsto \pmatrix{ 0 & 0 \cr 0 & u^{-1 \dagger} \cr} S
\pmatrix{0 & 0 \cr 0 & u^{-1} \cr}.
\eeq
We can then write the effective K\"ahler potential
\eq
\scr L_D = \myint\dtt \bigl[ \mybar A_+ A_+ + \mybar B_+ S B_+
+ \mybar B_- S^{-1 T} B_- \bigr],
\eeq
and the effective superpotential
\eq
\scr L_F = \myint\dt m_{\rm eff} B_+ B_- + \hc
\eeq
Note that different $U(N - 2)$ invariants are related in both the
superpotential and the K\"ahler potential.

The second example illustrates that the $\hat{H}$ invariance relates
invidual $H$-invariant terms in the superpotential. It has global $G =
U(N) \times U(1)_R$ symmetry with fields
\eq
\Phi_+ \sim (N; \sfrac 12), \qquad
\Phi_- \sim (\mybar N; \sfrac 12),
\eeq
where $U(1)_R$ is defined by
\eq
U(1)_R:\ \Phi_\pm(x, \theta) \mapsto e^{i\al / 2}
\Phi_{\pm}(x, \theta e^{i\al}).
\eeq
The most general superpotential compatible with these symmetries is
\eq
W = G (\Phi_+ \Phi_-)^2.
\eeq
This term can be imagined to arise from integrating out a heavy singlet
chiral superfield in a renormalizable theory.\footnote{$^*$}{For instance,
$W= \lambda \Phi_+ \Phi_- \chi + M \chi^2$, where $\chi$
has quantum numbers $(1;1)$. The superpotential including the matter
fields eq.~(3.65) can also be rewritten using additional singlet
fields, with our results remaining unchanged. The non-renormalizable
forms used in the text simplify some of the expressions.}
It is easy to check that there are no additional accidental symmetries.
There are \susc\ ground states for
\eq
\avg{\Phi_+} = \pmatrix{v_+ \cr 0 \cr \vdots \cr 0 \cr}, \qquad
\avg{\Phi_-} = \pmatrix{0 \cr \vdots \cr 0 \cr v_- \cr}.
\eeq
The potential is minimized for arbitrary $v_\pm$, so the potential has
2-dimensional space of flat directions.
For simplicity, we will analyze the theory for the special case $v_- = 0$.

The theory then has a symmetry breaking structure similar to that of the first
example in subsection 2.3.
There is an unbroken $U(1)_{R'}$ symmetry that is a combination of the
original $U(1)_R$ and a broken $U(N)$ generator:
\eq
U(1)_{R'}:\ \Phi_\pm(x, \theta) \mapsto e^{i\alpha / 2} e^{\pm i\alpha T}
\Phi_\pm(x, \theta e^{i\alpha}),
\eeq
where
\eq
T = \bordermatrix{& 1 & N - 1 \cr\ \ \ 1 & -\frac 12 & 0 \cr N - 1 & 0 & 0 \cr}
\in U(N).
\eeq
Therefore, the only effect of the $U(1)_R$ symmetry is the existence of the
symmetry $U(1)_{R'}$ in the \leff, and we write (for $v_- = 0$)
\eq
G = U(N), \qquad
H = U(N - 1).
\eeq
The group $\Hh$ is then the same as in eq.~\exHh.
There are $2N$ SNGB's in this model, which are conveniently parameterized by
\eq
\Phi_+ = \xi^{-1} \cdot \avg{\Phi_+},
\eeq
where
\eq
\xi = e^{i\Pi} \in \Gc / \Hh, \qquad \Pi =
\bordermatrix{ & 1 & N - 1 \cr
\ \ \ \ 1 & \sig & 0 \cr
N - 1 & \pi & 0 \cr}.
\eeq
There is a flat direction parametrized by the real part of $\sigma$,
consistent with the number of extra SNGB's as discussed in section 2.3.
There are also $N - 1$ light chiral matter fields
\eq
\label\oldmatter
\Xi_- \equiv
\bordermatrix{& 1 & N - 1 \cr\ \ \ \ 1 & 0 & 0 \cr N - 1 & 0 & 1 \cr}
\xi^T \Phi_- \equiv
\bordermatrix{&\cr \ \ \ \ 1 & 0 \cr N - 1 & C_- \cr}.
\eeq
There is no superpotential allowed for the light matter fields because of
$U(1)_{R'}$ symmetry.

To get a more interesting \leff, we again add additional fields.
We also impose an additional $U(1)$ symmetry, so that the full symmetry is
$U(N) \times U(1) \times U(1)_R$.
The fields are now
\eq
\eqalign{
&\Phi_+ \sim (N; 0, \sfrac 12), \qquad
\Phi_- \sim (\mybar N; 0, \sfrac 12), \cr
&\Sig_+ \sim (N; 1, \sfrac 12), \qquad
\Sig_- \sim(\mybar N; -1, \sfrac 12). \cr}
\eeq
The most general superpotential is
\eq
\label\allmatter
\eqalign{
W &= G_1 (\Phi_+ \Phi_-)^2
+ G_2 (\Sig_+ \Sig_-)^2 \cr
&\qquad + G_3 (\Phi_+ \Phi_-) (\Sig_+ \Sig_-)
+ G_4 (\Phi_+ \Sig_-) (\Sig_+ \Phi_-). \cr}
\eeq
This theory has a vacuum with $\avg{\Phi_\pm}$ as before (we again take
$v_- = 0$), and
\eq
\avg{\Sig_\pm} = 0,
\eeq
giving rise to the same symmetry-breaking pattern discussed above.
The low-energy matter fields are the fields $\Xi_-$ in eq.~\oldmatter, as
well as
\eq
\Psi_+ \equiv \xi^{-1} \cdot \Sig_+, \qquad
\Psi_- \equiv \xi^T \cdot \Sig_-.
\eeq
If we write
\eq
\Psi_\pm = \bordermatrix{&\cr \ \ \ \ 1 & A_\pm \cr N - 1 & B_\pm \cr},
\eeq
The effective superpotential is
\eq
W_{\rm eff} = G_{\rm eff} (\Psi_+ \Psi_-)^2
= G_{\rm eff} \left[ (A_+ A_-)^2 + 2 (A_+ A_-) (B_+ B_-) + (B_+ B_-)^2 \right].
\eeq
Just as in the example in subsection 3.2, there are three $H$ invariants
related
by $\Hh$.
Also, terms such as $A_+^4$ are allowed by $H$ as well as $U(1)_{R'}$, but are
forbidden by $\Hh$.
Terms proportional to powers of $A_-$ are allowed by $\Hh$ symmetry, but are
forbidden by the unbroken $U(1)$.

As described in the previous subsection, the effective K\"ahler potential for
the model is written in terms of fields transforming under $U(N - 1)$ \rep s.
In this example, the fields are
\eq
B_+ \mapsto u B_+, \qquad
A_- \mapsto A_-,
\eeq
and the non-holomorphic fields
\eq
\pmatrix{\twi A_+ \cr 0 \cr} \equiv \pmatrix{1 & 0 \cr 0 & 0 \cr}
e^{\scr W} \Psi_+, \qquad
\pmatrix{0 \cr \twi B_- \cr} \equiv \pmatrix{0 & 0 \cr 0 & 1 \cr}
e^{-\scr W} \Psi-,
\eeq
transforming as
\eq
\twi A_+ \mapsto \twi A_+, \qquad
\twi B_- \mapsto u^{-1} \twi B_-.
\eeq
The K\"ahler potential for the matter fields is simply the most general
$U(N - 1)$ invariant function of these fields.

The final example has a matter field whose mass term is forbidden by $\Hh$,
even though a mass is allowed by $H$ alone.
The model is a simple variation on the one just discussed:
the symmetry is $SU(N) \times U(1)_R$ with ``Higgs'' fields
\eq
\Phi_+ \sim (N; \sfrac 12), \qquad
\Phi_- \sim (\mybar{N}; \sfrac 12),
\eeq
and ``matter'' fields
\eq
\Sigma_+ \sim (N; \sfrac 32).
\eeq
In addition, we impose a $Z_2$ symmetry under which $\Phi_\pm$ is even and
$\Sigma_+$ is odd.
The superpotential is then simply
\eq
W = G (\Phi_+ \Phi_-)^2.
\eeq
The superpotential has an accidental $U(N)$ symmetry acting only on $\Sigma_+$,
but this can be broken in the K\"ahler potential by terms such as
\eq
\myint \dtt\,\, \mybar\Phi_+^4 \mybar\Phi_-^2\Sigma_+^2,
\eeq
where the indices are contracted in the obvious way.
In the effective theory, the light matter fields are
\eq
\Psi_+ \equiv \xi^{-1} \Sig_+ \equiv \pmatrix{A_\pm \cr B_\pm \cr}.
\eeq
A mass term for $A_+$ is allowed by $H$, but forbidden by $\hat H$, since
\eq
A_+ \mapsto A_+ + a \cdot B_+,
\eeq
where $a$ is defined in eq.~\exHh.

\section{Matter Fields: General Case}
In this section, we relax the assumption that the low-energy \leff\ arises
from a weakly-coupled theory, and explore the action of the group $\Hh$ on
the matter fields in a general \leff\ satisfying the assumptions stated in
subsection 2.2.
For the weakly-coupled case, we found that the $\Hh$ action on the matter
fields is linear, and that the $\Hh$ \rep s that arise can be embedded in
$\Gc$ \rep s.
We will show by explicit examples that the freedom to make field redefinitions
does not in general allow us to define matter fields on which $\Hh$ acts
linearly.
Furthermore, even if we restrict attention to linear $\Hh$ \rep s, we show that
they cannot be embedded in $\Gc$ \rep s in general.
This seems to make it impossible to write $\Gc$-invariant kinetic terms.
We therefore do not have a good understanding of the general \leff, and this
section is mainly an attempt to quantify our ignorance.

\subsection{Linearization}
We first show that we can redefine the matter fields so that the action of
$\Kc$ is linear.
Expanding the transformation eq.~\Psitrans\ for small $\Psi$, we have
\eq
\Psi \mapsto R(\hat{h}(g, \xi)) \Psi + O(\Psi^2).
\eeq
Note that there is no $\Psi$-independent term on the \rhs\ because
$T(\hat{h})(0) = 0$.
Following ref.~\CWZ, we then define
\eq
\Xi \equiv \int_K \omega(k)\, R(k)^{-1} T(k)(\Psi),
\eeq
where the integral is over the compact subgroup $K \subset \Kc$, and
$\omega(k)$ is the invariant group measure on $K$.
Despite the fact that the integral is defined only over $K$, the fields
$\Xi$ actually transform linearly under all of $\Kc$.
To see this, note that under $\ell \in \Kc$,
\eqa
\Xi \mapsto& \int_K \omega(k)\, R(k^{-1}) T(k)(T(\ell)(\Psi)) \eolnn
&= \int_K \omega(k)\, R(k)^{-1} T(k\ell)(\Psi) \eolnn
&= R(\ell) \int_{K\ell} \omega(k')\, R(k')^{-1} T(k')(\Psi), \eeol
\eeq
where we have changed variables to $k' = k\ell$ in the last line, so the
integration is now over $K\ell \equiv \{ k\ell \;|\; k \in K \}$.
Since the group action $T(k')(\Psi)$ and the group measure are holomorphic in
the group parameters, we can deform the contour back to $K$, and
obtain\footnote{$^*$}{The measure $\omega(k)$ is the natural
analytic continuation of the Haar measure on $K$ to $\Kc$.
Note that $\omega(k)$ is closed in $\Kc$: It is
holomorphic in the group coordinates, $\bar{\partial} \omega(k) = 0$, and
also the highest form in the holomorphic coordinates, $\partial
\omega(k) = 0$. Therefore, one can continuously deform the integration
region within $\Kc$ as long as one does not encounter singularities in the
integrand.}
\eq
\Xi \mapsto R(\ell) \cdot \Xi, \qquad \ell \in \Kc.
\eeq

The argument above relies crucially on the fact that $K$ is compact, since
the group-invariant measure is not defined for general non-compact groups.
To see that this is not just a technicality, we give an explicit example of
a $\Hh$ group action that cannot be linearized by any redefinition of the
matter fields that preserves the origin.
Consider a case with $G = SU(2)$ broken by an order parameter transforming
in the defining \rep.
In this case, we can make an $SU(2)$ \trans\ to put the order parameter in
the form
\eq
\avg\Phi = \pmatrix{v \cr 0 \cr},
\eeq
and we see that $H = 1$.
The group $\Hh$ is given by the set of $2 \times 2$ matrices of the form
\eq
\hat h = \pmatrix{1 & a \cr 0 & 1 \cr}
\eeq
with $a$ complex.
It is easy to see that the $a$'s add under group multiplication, so $\Hh$ is
isomorphic to the group of translations in the complex plane.
Now consider a single matter chiral superfield $\Psi$ transforming as
\eq
\label\badtrans
\Psi \mapsto \frac{\Psi}{1 + a \Psi}.
\eeq
This transformation leaves the origin invariant, and it is easily checked
that it satisfies the group multiplication law.
We wish to define new matter fields $\Xi(\Psi)$ that transform linearly
under $\Hh$.
These fields should transform as
\eq
\Xi(\Psi) \mapsto \Xi(\Psi - a\Psi^2 + O(a^2)) = \Xi + at\Xi + O(a^2),
\eeq
where $t$ is the ``generator'' in the linearized \trans.
Equating the $O(a)$ terms gives the requirement
\eq
-\Psi^2 \frac{d\Xi}{d\Psi} = t\Xi.
\eeq
The general solution is
\eq
\Xi = C e^{t / \Psi},
\eeq
which does not satisfy the condition that $\Xi = 0$ when $\Psi = 0$.
Thus we see that the \trans\ eq.~\badtrans\ cannot be linearized.

To see what $\Hh$ invariants we can construct in this example, note that
eq.~\badtrans\ can be rewritten as
\eq
\label\funnytrans
\frac 1\Psi \mapsto \frac 1\Psi + a.
\eeq
Therefore, we can write $\Hh$ invariant terms such as
\eq
\myint \dtt \frac{1}{\mybar\Psi \Psi}.
\eeq
(This shifts by a total derivative under the \trans\ eq.~\funnytrans.)
This can perhaps be interpreted to give a sensible effective field theory
by expanding around $\Psi = \infty$, but the resulting \leff\ can certainly
not be interpreted as describing fluctuations about $\Psi = 0$.
We would therefore be inclined to regard this \leff\ as ``unphysical.''
We do not know whether the features found in this example are general to all
non-linearizable group actions.

\subsection{Non-embeddable Representations}
We now restrict attention to \leff s where the $\Hh$ action on the matter
fields can be linearized, and make some brief comments on general $\Hh$
\rep s.
We point out that there are simple $\Hh$ \rep s that cannot be embedded into
$\Gc$, and that there appears to be no way to write kinetic terms for fields
transforming according to these \rep s.

As discussed in the Appendix, the general \rep s of $\Hh$ contain 1-dimensional
\rep s (characters) of $N$.
If these characters are non-trivial, then the \rep\ cannot be embedded into
a $\Gc$ \rep, since elements of the subgroup $N \in \Gc$ are represented by
unipotent matrices in a $\Gc$ \rep.
We can use this fact to construct simple $\Hh$ \rep s that cannot be
embedded in $\Gc$ \rep s.

A simple example is obtained by considering again the symmetry breaking
pattern $SU(2) \to 1$ discussed in the previous subsection.
Consider now a field $\Psi$ transforming under $\Hh$ as
\eq
\Psi \mapsto e^{t a} \Psi
\eeq
for some constant $t$.
By the arguments in the Appendix, this \rep\ cannot be embedded in $\Gc$.
The importance of this is that we do not know any way to couple the gauge field
spurion $\scr V$ to $\Psi$.
(In subsection 3.4, we saw that couplings of $\scr V$ are crucial for writing
$\Gc$-invariant kinetic terms for the embeddable $\Hh$ \rep s.)
In the present case,
\eq
\mybar\Psi \Psi \mapsto e^{ta + \bar t \bar a} \mybar\Psi \Psi,
\eeq
and there appears to be no way to use $\scr V$ to construct an
$SU(2)^{\rm c}$-invariant kinetic term.

We do not know whether there are any non-embeddable $\Hh$ \rep s for which
one can write a sensible \leff.
The question is an interesting one, since such matter fields would be analogs
of states with fractional charge, such as dyons.

\section{Conclusions}

In this paper we have discussed the structure of \susc\ \leff s describing the
low-energy physics in a situation where a global symmetry group $G$ is
spontaneously broken down to a subgroup $H$ while supersymmetry remains
unbroken.
This \leff\ contains fields describing the supersymmetric Nambu--Goldstone
bosons (SNGB's), as well as possible additional light ``matter'' fields.
By introducing external ``spurion'' gauge fields for $G$, the symmetry is
formally enhanced to $\Gc$, the complexification of $G$.
By studying the way in which this external gauge field can appear in the \leff,
we have shown that the effective couplings of the matter fields are constrained
by the group $\Hh$, the largest unbroken subgroup of $\Gc$.
The structure of $\Hh$ is rather non-trivial: it can be decomposed into a
semidirect product $\Kc \wedge N$, where $K$ is compact and $N$ is unipotent.
$K$ contains $H$, but $K$ is larger than $H$ in general.

We have shown how to write a manifestly \susc\ \leff\ for the SNGB's, but our
main results concern the matter fields.
We showed that the superpotential for the matter fields is invariant under
$\Hh$.
In cases where $\Hh$ can be larger than $\Hc$, the coefficients of
$H$-invariant terms therefore obey relations imposed by $\Hh$ invariance.
The K\"ahler potential for the matter fields is determined by the most general
$K$-invariant function of the matter fields, with the explicit breaking down to
$H$ determined as a function of the order parameter.
Both these results are considerably stronger than the simple $H$-invariance
one naively expects.

The assumptions made in deriving these result are that the holomorphy of the
group action is preserved in the quantum theory, and that the action of
$\hat{H}$ on the matter fields can be taken to be a linear \rep\ embedded in a
$G$ \rep;
both of these assumptions are valid in weakly-coupled theories.
Relaxing these assumptions, we show that there are $\Hh$ actions on the matter
fields that cannot be made linear by field redefinitions, and there are $\Hh$
\rep s for which it appears to be impossible to write a $\Gc$-invariant kinetic
term.
It is not clear to us whether a physically sensible \leff\ can be constructed
from matter fields transforming under these more general $\Hh$ actions.

\section{Acknowledgments}
We would like to thank S. Coleman, S. Thomas, P. Watts, and B. Zumino
for discussions, and especially D. Vogan for much help with group theory.
MAL is supported in part by DOE contract DE-AC02-76ER03069 and by NSF
grant PHY89-04035.
JMR wishes to thank ORISE for the support of a DOE Distinguished
Postdoctoral Research Fellowship in the early stages of this work at
LBL, and DOE contract DE-FG02-90ER40524 and the W.~M.~Keck Foundation
for support at the IAS.
This work was supported by the Director, Office of Energy Research,
Office of High Energy and Nuclear Physics, Division of High Energy
Physics of the U.S. Department of Energy under Contract DE-AC03-76SF00098.

\appendix{A}{Structure of $\Hh$ Representations}
In this appendix, we prove the structure theorem alluded to in subsection
3.2.\footnote{$^*$}{We thank D. Vogan for this argument.}
Given any \rep\ $R$ of $\Hh$, Engel's theorem tells us that there is a
basis in which
\eq
R(n) = \pmatrix{\lam_1(n) S_1(n) & 0 & \cdots & 0 \cr
0 & \lam_2(n) S_2(n) & 0 & \vdots \cr
\vdots & 0 & \ddots & 0 \cr
0 & \cdots & 0 & \lam_r(n) S_r(n) \cr},
\eeq
for $n \in N$.
Here, $\lam_1, \ldots, \lam_r$ are 1-dimensional \rep s (characters) of $N$,
and $S_1, \ldots, S_r$ are unipotent matrices:
that is, they are upper-triangular with $1$'s on the diagonal.
However, if $R$ is a \rep\ of $\Hh$ obtained by reducing a \rep\ of $\Gc$,
then elements of $N$ are represented by matrices with
$\lam_1, \ldots, \lam_r \equiv 1$.
One way to see this is to note that the \rep s of $\Gc$ can be obtained by
taking tensor products of fundamental \rep s and reducing them, and these
operations preserve the property of having 1 as an eigenvalue.
Therefore, every unipotent element of $\Gc$ will be represented by a unipotent
matrix.

We now restrict attention to the case where the $\Hh$ \rep\ is embedded in a
$\Gc$ \rep.
In that case, we denote the state space for the \rep\ $R$ by $V$ and define
the subspace
\eq
V_1 \equiv \{ v \in V \;|\; R(n) v = v {\rm\ for\ all\ } n \in N \}.
\eeq
The considerations above tell us that $V_1 \ne 0$.
It is also easy to see that $V_1$ is invariant under $\Kc$, since for all
$v \in V_1$ we have
\eq
R(n) \cdot R(k) v = R(k) R(k^{-1} n k) v = R(k) v
\eeq
for all $k \in \Kc$, $n \in N$ (because $N$ is a normal subgroup of $\Hh$).
This means that there is a basis for $V$ in which the \rep\ matrices have the
block form
\eq
R(k n) = \pmatrix{ R_1(k) & * \cr 0 & * \cr}
\pmatrix{ 1 & * \cr 0 & * \cr}
= \pmatrix{ R_1(k) & * \cr 0 & * \cr}.
\eeq
It is easy to see that the block in the lower-right corner is again a \rep\
of $\Hh$, and we can apply the same argument to it.
Therefore, we obtain that any embedded $\Hh$ \rep\ is equivalent to a \rep\ of
the form given in eq.~\Hhrepform\ in the main text.

It is interesting that a ``folk theorem'' in the mathematics community states
that the converse of this result is also true:
any $\Hh$ \rep\ of the form eq.~\Hhrepform\ is isomorphic to a sub\rep\ of a
$\Hh$ \rep\ obtained by restricting a $\Gc$ \rep\
\ref\Vogan{D. Vogan, private communication.}.


\listrefs
\vfill\eject
\bye